**The contribution of local variations in hue or contrast to *"symmetry of things in a thing"***

Birgitta Dresp-Langley

Centre National de la Recherche Scientifique (CNRS)
UMR 7357 ICube Lab Strasbourg University
Strasbourg, France

e-mail: birgitta.dresp@unistra.fr

**Abstract**

Symmetry contributes to processes of perceptual organization in biological vision and influences the quality and time of goal directed decision making in animals and humans, as discussed in recent work on the examples of *'symmetry of things in a thing'* and bilateral shape symmetry. The present study was designed to show that selective chromatic variations in geometric shape configurations with mirror symmetry can be exploited to highlight functional properties of *'symmetry of things in a thing'* in human vision. The experimental procedure uses a psychophysical two-alternative forced choice technique, where human observers have to decide as swiftly as possible whether two shapes presented simultaneously on a computer screen are symmetrical or not. The stimuli are computer generated 2D shape configurations consisting of multiple elements, with and without systematic variations in local color, color saturation, or achromatic contrast to manipulate *'symmetry of things in a thing'*. All stimulus pairs presented had perfect geometric mirror symmetry. The results show that varying the color of local shape elements selectively in multi-chromatic and mono-chromatic shapes significantly slows down perceptual response times, which are a direct measure of uncertainty. It is concluded that local variations in hue or contrast may produce functionally important variations in *'symmetry of things in thing'*, which is revealed here as a relevant perceptual variable in symmetry detection. Disturbance of the latter increases stimulus uncertainty, and thereby affects the perceptual salience of mirror symmetry in the time course for goal-relevant human decision.

*<u>Keywords</u>: biological vision; bilateral symmetry; geometry; color; contrast; symmetry of things in a thing.*

**Background**

Vertical mirror symmetry is a particularly salient form of visual symmetry, processed at early stages in human vision and producing greater or lesser detection reliability [1-12] depending on local features of the stimulus display with greater or lesser stimulus certainty. Shape symmetry is a visual property that attracts attention and determines the perceived volume or perceptual salience of objects in the two-dimensional image plane [13-18]. Aesthetic judgments and choice preferences are also influenced by symmetry [19-21]. Symmetry is detected not only by human but also by other species, such as insects, for example [22]. In the domain of visual objects, symmetry plays an important role in conceptual processes for structural design, and is abundantly exploited by engineers and architects. As examples for 'symmetry of things in a thing', we may refer to the perceived symmetry of curves, which may be seen as single things or as multiples of one and the same thing in a complex shape or object, or to that of two-dimensional fractal trees based on geometrical principles producing *'symmetry of things in a thing'* inspired by nature [22-26], as illustrated in the examples here below (Figs.1, 2):

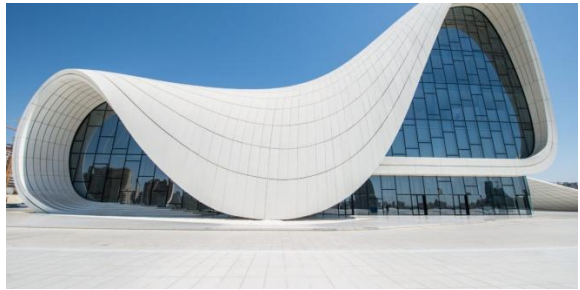

Figure 1: Free-form structure designed by Zaha Hadid, Pritzker Prize 2004. Curve symmetry embedded in structural complexity is frequently exploited in contemporary architecture and harks back to vernacular architecture inspired by symmetry in nature.

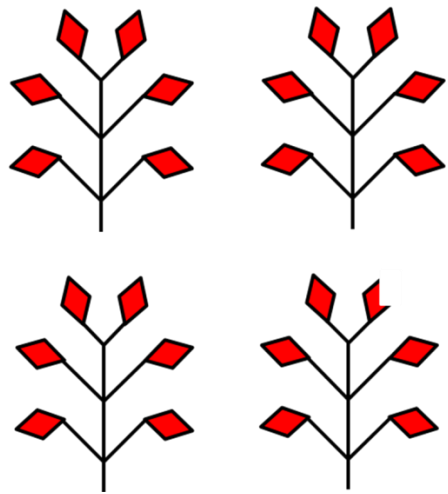

Figure 2: Bilateral *"symmetry of things in thing"* in 2D fractal trees (top) inspired by nature [22]. The smallest fractal change in one of the trees perturbs the complex geometry that produces *"symmetry of things in a thing"*.

In this study, pairs of abstract shapes with perfect vertical mirror symmetry, with computer generated disturbances of *'symmetry of things in a things'* by selectively introducing local variations in the hue or contrast of single shape elements. How these disturbances influence human response times to symmetry is investigated.

**Materials and methods**

All image configurations were computer generated and displayed on a high resolution color monitor (EIZO COLOR EDGE CG 275W, 2560x1440 pixel resolution) connected to a DELL computer equipped with a high performance graphics card (NVIDIA). Color and luminance calibration of the RGB channels of the monitor was performed using the appropriate Color Navigator self-calibration software, which was delivered with the screen and runs under Windows 7. RGB values here correspond to ADOBE RGB. All luminance levels were cross-checked with an external photometer (OPTICAL, Cambridge Research Systems). RGB coordinates, luminance parameters (cd/m$^2$), and color coordinates of the shape elements in the images are given in Table 1.

Table 1. Local color parameters

| Colors | Hue | Sat | Lightness | R-G-B |
|---|---|---|---|---|
| BLUE | 240 | 100 | 50 | 0-0-255 |
| RED | 0 | 100 | 50 | 255-0-0 |
| GREEN | 120 | 100 | 50 | 0-255-0 |
| MAGENTA | 300 | 100 | 50 | 255-0-255 |
| YELLOW | 60 | 100 | 50 | |
| | | | | |
| BLUE | 180 | 95 | 50 | 10-250-250 |
| RED | 0 | 100 | 87 | 255-190-190 |
| GREEN | 120 | 100 | 87 | 190-255-190 |
| MAGENTA | 300 | 25 | 87 | 255-190-255 |
| YELLOW | 600 | 65 | 67 | 255-255-190 |

All shape elements had identical size, and all shapes identical numbers of "fractal" elements. Two mirror shapes in each image always had perfect geometrical mirror symmetry (bilateral symmetry). Some of the images are shown, for illustration, in Figure 1. The position (left, right) of local color/contrast variations in the shape pairs was counterbalanced between chromatic and achromatic displays as shown here below (Fig.3 ), and also within each level of this factor yielding a total of 20 different images. These were presented in a random order in two successive experimental sessions per individual. The subject pool consisted of mostly students, with normal or corrected-to-normal vision. All of them were naïve to the purpose of the experiment and run in individual sessions. They were comfortably seated in a semi-dark room, in front of the EIZO monitor at a viewing distance of about one meter. Each individual received the same standard instructions for the psychophysical task, which was to "decide as quickly as possible whether two shapes in an image were symmetrical or not". To this effect one of two keys on the computer keyboard had to be pressed: '1' for 'yes'; '2' for 'no'. The individual response time (RT) computed by the CPU corresponds to the time between the onset of an image and the time a response key was pressed.

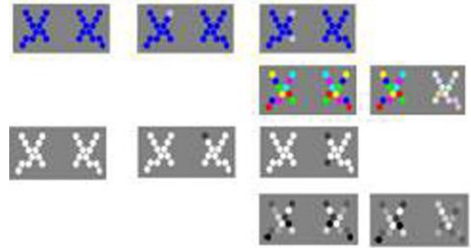

Figure 3: Shape pairs with perfect mirror symmetry. Local variations in color or contrast were selectively manipulated to vary *"symmetry of things in a thing"*. The spatial position (left/right shape in a pair) of the local variations was counterbalanced between chromatic and achromatic displays as shown here for illustration, and also within each level of this factor.

The response times (RT) were written to an Excel file with labeled columns for each subject, session, and trial. For 20 images per individual session, two repeated sessions per participant, and 15 participants, a total of 600 choice response time data were generated.

**Results and discussion**

The average response time data per subject and factor level were submitted to descriptive and statistical analyses for to a Cartesian analysis plan *Complexity$_5$ x Appearance$_2$ x* 15*,* with five levels of the '*Complexity'* factor for different degrees of local hue/contrast manipulation of shape elements (Fig. 3), two levels of the *'Appearance'* factor for achromatic *versus* chromatic visual appearance of the stimuli (Fig. 3), and 15 individual average response times (RT) for each factor level yielding N=150 data for the statistical analysis (2-Way ANOVA) with a total number of N-1=149 degrees of freedom (DF). The results from this analysis are shown here below in Table 2.

Table 2. Result of the statistical analysis (2-Way ANOVA)

| Factor | DF | F | P |
|---|---|---|---|
| COMPLEXITY | 4 | 1053.3 | <.001 |
| APPEARANCE | 1 | 92.4 | <.001 |
| INTERACTION | 4 | 126.6 | <.001 |

For a graphic representation, average RT for different levels of complexity in chromatic and achromatic stimuli were averaged over complexity factor levels with the least differences, separately for each of the two levels of the color (hue, appearance) factor, a shown in the histogram here below (Fig. 4).

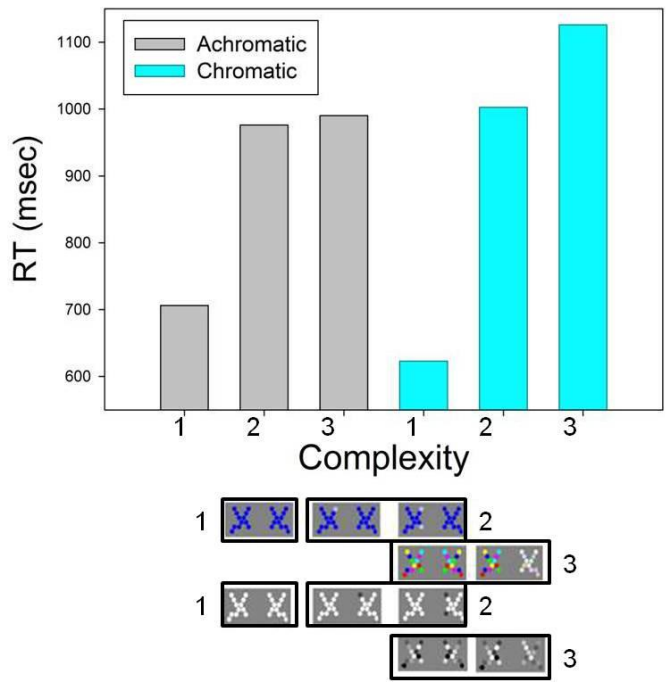

Figure 4: Average RT as a function of the three major levels of complexity in the achromatic (left) and chromatic (right) stimuli.

The results represented here above (Fig. 4) show that when complexity is the least (1), achromatic shape pairs produce the longest RT for symmetry detection. When complexity is maximal (3), the opposite is observed with chromatic shape pairs producing the longest RT for symmetry detection. This is consistently reflected in the significant statistical interaction (Table 2) between the complexity and the color (hue, visual appearance) factors. Average RT data for the five levels of the complexity factor in each level of the color (hue, visual appearance) factor were submitted to linear regression analysis, and the results plotted in separate graphs for each level of the color (hue, appearance) factor. The results from the linear regression analysis with correlation coefficients $R$ are shown here below (Fig. 5).

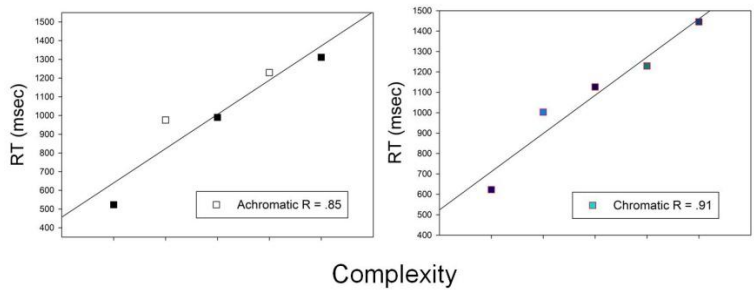

Figure 5: Average RT as a function of the five level complexity in the achromatic (left) and chromatic (right) displays. RT is shown to increase linearly with stimulus complexity. The correlations coefficients $R$ confirm the goodness of the linear fits.

The results represented here above (Fig. 5) show that RT increases linearly with the complexity of the displays. An interpretation of this finding in terms of a direct consequence of Information Theory and Hick's Law [27], which presumes a direct relationship between RT and sensory system uncertainty (SU) where RT increases linearly with amount of transmitted information/stimulus uncertainty, is suggested. Hick's Law is represented graphically here below (Fig. 6).

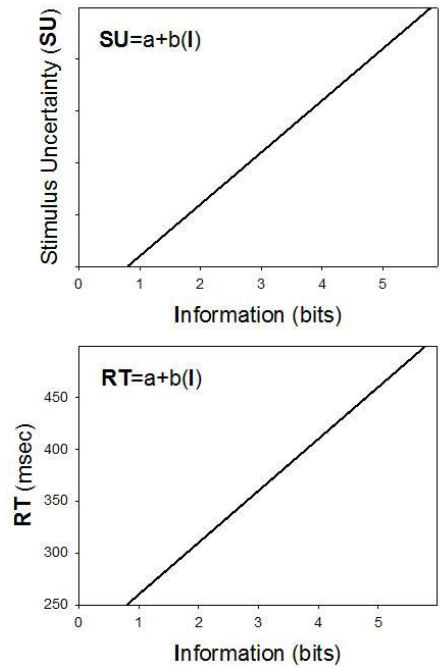

## Conclusions

The results highlight that local color (hue) or contrast variations disturb the *"symmetry of things in a thing"* in mathematically symmetrical stimuli. This observation can be interpreted in terms of increasing perceptual complexity with increasing local color variations. Increased perceptual complexity increases stimulus uncertainty, as consistently reflected by the observed increase in RT, which can be interpreted as the direct consequence of Hick's Law (1955) and information theory.